\documentclass[aps,prb,twocolumn,amsmath,amssymb,superscriptaddress,floatfix,longbibliography, nofootinbib]{revtex4-2}

\usepackage{amsmath}
\usepackage{amssymb}
\usepackage{amsthm}
\usepackage{mathtools}
\usepackage{mathdots}
\usepackage{amsfonts}
\usepackage{bbm}
\usepackage{xr}
\usepackage{bbold}
\usepackage{epsfig,color,dsfont,upgreek,physics}
\usepackage{mathrsfs}
\usepackage{multirow}
\renewcommand{\selectlanguage}[1]{}

\usepackage{graphicx}
\usepackage{dcolumn} 
\usepackage{bm} 
\usepackage{float} 
\usepackage[driverfallback=dvipdfm]{hyperref} 
\usepackage{longtable}
\usepackage{ulem}   
\allowdisplaybreaks

\renewcommand\[{\begin{equation}}
\renewcommand\]{\end{equation}}

\begin{document}

\title{Superconducting Diode Effect in Multiphase Superconductors}

\author{Daniel Shaffer}
\affiliation
{Department of Physics, University of Wisconsin-Madison, Madison, Wisconsin 53706, USA}

\author{Dmitry V. Chichinadze}
\affiliation
{National High Magnetic Field Laboratory, Tallahassee, Florida 32310, USA}

\author{Alex Levchenko}
\affiliation
{Department of Physics, University of Wisconsin-Madison, Madison, Wisconsin 53706, USA}

\begin{abstract}
We identify a particular mechanism for the intrinsic superconducting diode effect (SDE) in multiphase superconductors. Using a Ginzburg-Landau and a microscopic two-band model, we find phase transitions into a mixed phase with finite-momentum Cooper pairs and SDE with high (including maximal) diode efficiencies, despite the individual phases exhibiting no SDE and equal inversion parity. We thus show that parity mixing -- invoked in previous proposals -- is not a crucial ingredient for SDE. The new mechanism may be relevant in a multitude of known multiphase superconductors like UTe\(_2\).
\end{abstract}
\date{\today}
\maketitle


\section{Introduction}

Barring early observations from boundary effects \cite{SwartzHart67, CerbuVondel13} and theoretical predictions \cite{LevitovNazarovEliashberg85, GeshkenbelhLarkin86, edelstein_ginzburg_1996, VodolazovPeeters05}, the superconducting diode effect (SDE) has only recently become a subject of much interest, spurred by its experimental discovery in bulk materials and Josephson junctions
\cite{AndoYanaseOno20,  LinScheurerLi22, BauriedlParadiso22, HouMoodera23,  ZhangChichinadzeFuLi24, LeLin24, AsabaYanaseMatsuda24} and subsequent improved theoretical understanding \cite{DaidoYanase22, YuanFu22, IlicBergeret22, KapustinRadzihovsky22, HasanLevchenko23, HasanShafferKhodasLevchenko24, OsinLevchenkoKhodas24, GaggioliGeshkenbein24, YerinGiazotto24}; 
for reviews see \cite{NadeemFuhrerWang23, AmundsenLinderZuticBanerjee24}.
Defined as the nonreciprocity of critical currents in bulk superconductors, SDE has many potential applications in superconducting and quantum electronics.
There is consequently a high demand for novel realizations of SDE, with recent proposals including chiral SDE \cite{NunchotYanase24}, unidirectional SDE \cite{DaidoYanase23}, SDE due to magnetization gradients \cite{KotetesAndersen23}, SDE due to spontaneous symmetry breaking and coexisting non-SC order \cite{BanerjeeScheurer24}, and SDE in altermagnets \cite{BanerjeeScheurer24alt, ZhangNeupert24, ChakrabortyBlackSchaffer24}.

All proposals necessarily require the explicit or spontaneous breaking of both inversion \(\mathcal{I}\) and time reversal \(\mathcal{T}\) symmetries. The crystallographic symmetry imposes additional constrains on the Lifshitz invariants responsible for the finite SDE. The Lifshitz invariants for different crystallographic point groups are tabulated in Ref. \cite{SmidmanAgterberg17}. Breaking of $\mathcal{I}$ and $\mathcal{T}$ symmetries   
is generically accompanied by finite-momentum pairing similar to that in the FFLO state \cite{LO,FF, SamokhinTruong17} or \(\mathcal{T}\)-breaking pair-density waves (PDW) 
\cite{AgterbergReview20, WuThomaleRaghu22, ShafferBurnellFernandes23, ShafferSantos23, CastroShafferWuSantos23,  LiuHan24,  ZhangNeupert24}.
The finite-momentum pairing is strictly speaking neither necessary nor sufficient for SDE to occur, though it breaks the same symmetries. This is especially true for weak-coupling PDWs that take place due to Fermi surface nesting, with the Fermi surface symmetric under \(\mathbf{p+q}\rightarrow-\mathbf{p+q}\): that same symmetry, if exact, precludes the SDE.

The FFLO as a mechanism for SDE was considered in Ref. \cite{YuanFu22}. Similarly, SDE was initially theorised \cite{edelstein_ginzburg_1996, DaidoYanase22, YuanFu22} and reported \cite{AndoYanaseOno20, BauriedlParadiso22} in finite-momentum carrying helical SCs \cite{MineevSamokhin94, BauerSigrist12, SmidmanAgterberg17}, i.e. 
noncentrosymmetric SCs with spin-orbit coupling (SOC) placed in in-plane magnetic fields. Due to the broken \(\mathcal{I}\), these systems also in general exhibit parity mixing, e.g. \(s\)/\(p\)-wave or singlet/triplet mixing SC 
\cite{MockliKhodas19, ShafferKangBurnellFernandes20, WickramaratneAgterbergMazin20, HamillPribiag21, HaimLevchenkoKhodas22, ShafferBurnellFernandes23}. 
On the other hand, mixed-parity SC has also been considered in multiphase SCs near phase boundaries between SCs of opposite parity in systems without an explicit \(\mathcal{I}\)-breaking (with or without \(\mathcal{T}\)-breaking). In that case \(\mathcal{I}\) is instead spontaneously broken in the mixed or coexistence phase
\cite{Sergienko04, HinojosaFernandesChubukov14, YuxuanFu17, YangWu20}. In particular, a phase transition between SCs of opposite parity has been reported in CeRh\(_2\)As\(_2\) under an applied magnetic field \cite{KhimAgterbergHassinger21, NogakiYanase21, MockliRamires21, SzaboSigrist24}; another historical example is \(^3\)He \cite{Wheatley75, Leggett04}.
It has recently been shown \cite{ChazonoYanase23}  that SDE can indeed be realized in such coexisting mixed-parity SCs in the particular case of anapole SCs (which preserve \(\mathcal{IT}\)) \cite{KanasugiYanase22}, theoretically considered in UTe\(_2\) \cite{IshizukaYanase21, HakunoYanase24}.

In this work, we show that parity mixing is not essential for producing the SDE, and thus establish a more general mechanism for realizing SDE in multiphase superconductors.
This greatly extends the list of materials with potential for SDE, including \(s+d\) \cite{KhodasChubukov12,  FernandesChubukov16} and \(s+is\) \cite{MaitiChubukov13, SpeightBabaev21} SC proposed in iron-based high-T\(_c\) SCs and moir\'{e} SCs like twisted bilayer graphene \cite{ChichinadzeClassenChubukov20}.
The key theoretical observation behind the proposed mechanism is that even if the two nearby SC order parameters are uniform (have zero momentum pairing) and have equal parity, finite-momentum pairing with momentum \(q\) can still emerge in the coexistence phase.

\begin{figure*}[t]
\centering
\includegraphics[width=0.99\textwidth]{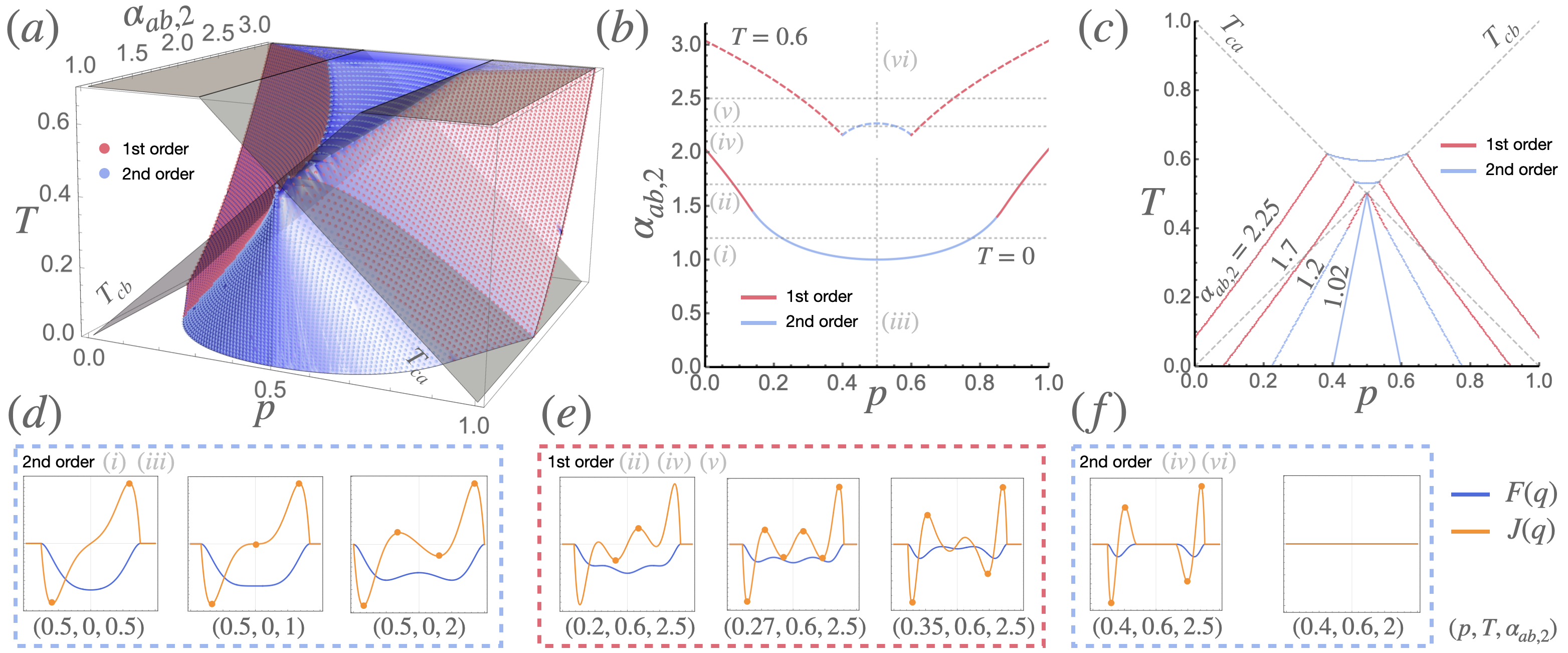}
\caption{(a) Phase diagram of free energy Eq. \ref{EqGL} with parameters \(T\), \(p\) and \(\alpha_{ab,2}\). Blue surface indicates the phase boundary of the non-uniform finite-momentum phase exhibiting SDE. Uniform (zero-momentum) SC boundaries for both \(a\) and \(b\) phases are shown in gray. Red and blue points indicate numerically computed first and second order phase transitions, respectively. (b) and (c) show representative cuts of the phase diagram at fixed \(T\) and \(\alpha_{ab,2}\), respectively, with first and second order phase transitions shown in red and blue. (d), (e), (f) show the schematic evolution of the condensation energy \(F\) (blue) and the supercurrent \(J\) (orange) vs Cooper pair momentum \(q\) across the lower-temperature second order, first order, and higher-temperature second order transitions, respectively (computed at various points \((p,T,\alpha_{ab,2})\) in the phase diagram indicated on bottom). All plots are in arbitrary units and centered at zero. Marked maxima (minima) of \(J\) correspond to the critical supercurrents \(J_{c+}\) (\(J_{c-}\)).}
\label{fig:PhaseDiagram}
\end{figure*}

We demonstrate this using a minimal Ginzburg-Landau (GL) free energy with two order parameters \(\Delta_a\) and \(\Delta_b\) with critical temperatures that can be tuned by some external parameter \(p\) (pressure, magnetic or electric field, doping, etc.). Crucially, a mixing term of the form \(\alpha_{ab}(q)\Delta_a^*\Delta_b+c.c.\) with \(\alpha_{ab}\propto q^2\) is allowed by symmetry. A first or second order phase transition then takes place into the nonuniform phase depending on the strength of \(\alpha_{ab}\), as shown in Fig. \ref{fig:PhaseDiagram}, spontaneously breaking both \(\mathcal{I}\) and \(\mathcal{T}\) symmetries (unless \(p\) is itself a \(\mathcal{T}\)-breaking field). High (including perfect) diode efficiencies in the vicinity of the phase transitions are found. We confirm the plausibility of the proposed mechanism by calculating the GL coefficients in a microscopic two-band model.


\section{Ginzburg-Landau Model}

We consider the simplest possible Ginzburg-Landau theory of two coexisting superconducting orders \(\Delta_a\) and \(\Delta_b\) (working in momentum space and assuming plane wave \(e^{i\mathbf{q\cdot r}}\) solutions):
\begin{align}
    \mathcal{F}&=\alpha_{a}(q) |\Delta_a|^2+(\alpha_{ab}(q) \Delta_a^*\Delta_b+c.c.)+\alpha_{b}(q) |\Delta_b|^2+\nonumber\\
    &+\beta_a|\Delta_a|^4+\beta_b|\Delta_b|^4\,,
\end{align}
where \(\Delta_a\) and \(\Delta_b\) are distinguished by their symmetry properties (i.e. they belong to different irreducible representations, e.g. they may be \(s\)- and \(d\)-wave, etc.) and we assume inversion symmetry. For simplicity we neglect the momentum dependence of fourth order coefficients, and assume both \(\Delta_a\) and \(\Delta_b\) carry the same momentum in the coexistence phases.  We furthermore neglected several symmetry-allowed quartic terms, which in principle can have important effects, especially on the complex phase between \(\Delta_a\) and \(\Delta_b\), but which we found not to qualitatively alter our main results (see Appendix \ref{AppA}--\ref{AppC}). The mixed coefficient \(\alpha_{ab}(\mathbf{q})\) has to vanish at \(\mathbf{q}=0\) and is either odd or even in \(\mathbf{q}\) if \(\Delta_a\) and \(\Delta_b\) have opposite or equal parity under inversion. The former case includes the anapole SC \cite{KanasugiYanase22} as a special case.

Here we consider the equal parity case and work to fourth order in \(q\), taking \(\alpha_{ab}(\mathbf{q})=\alpha_{ab,2}(q_x^2-q_y^2)/2=\alpha_{ab2}q^2 \cos \left(2\theta\right) /2\) where \(\theta\) is the momentum polar angle (note that by symmetry, it cannot be constant in \(\theta\)). Minimizing with respect to \(\theta\) and the relative phase between \(\Delta_a\) and \(\Delta_b\), we obtain
\begin{align}
    \mathcal{F}&=\alpha_{a}(q) |\Delta_a|^2-\alpha_{ab,2}q^2 |\Delta_a||\Delta_b|+\alpha_{b}(q) |\Delta_b|^2+\nonumber\\
    &+\beta_a|\Delta_a|^4+\beta_b|\Delta_b|^4
\end{align}
with \(\alpha_{f}(q)=\alpha_{f,0}+\alpha_{f,2}q^2+\alpha_{f,4}q^4\) for \(f=a,b\), and where \(\alpha_{ab,2}\) is now redefined to be real and positive. We also take \(\alpha_{f,0}(T,p)=A (T - T_{cf}(p))\) where \(T\) is temperature and \(p\) is some tuning parameter (modelling e.g. pressure, magnetic field, doping, etc.) that tunes between the \(a\) and \(b\) phases; for simplicity we take \(T_{ca}=1-p\) and \(T_{cb}=p\).

The mixed GL coefficient \(\alpha_{ab,2}\) is the new element of the model responsible for the field-free SDE. To understand this mechanism, we consider the GL saddle-point equations
\begin{align}\label{EqGL}
    \frac{\delta\mathcal{F}}{\delta|\Delta_f|}&=2\alpha_f(q)|\Delta_f|-\alpha_{ab,2}q^2|\Delta_{\bar{f}}|+4\beta_f|\Delta_f|^3=0\nonumber\\
    \frac{\delta\mathcal{F}}{\delta q^2}&=(\alpha_{a,2}|\Delta_a|^2-\alpha_{ab,2}|\Delta_a||\Delta_b|+\alpha_{b,2}|\Delta_b|^2)+\nonumber\\
    &+2\left(\alpha_{a,4}|\Delta_a|^2+\alpha_{b,4}|\Delta_b|^2\right)q^2=0\,,
\end{align}
where \(\bar{f}=a,b\neq f\).
The SDE occurs when \(q\neq0\) in the ground state (i.e. at the lowest energy saddle point). In a pure state (\(\Delta_a\neq 0\) and \(\Delta_b=0\), or vice versa), this is only possible if \(\alpha_{f,2}<0\) for either \(f=a,b\), which happens in the FFLO mechanism \cite{FF,LO,SamokhinTruong17}. Assuming no FFLO physics, i.e. \(\alpha_{f,2}>0\), it is still possible to have \(q=\pm q_0\neq0\) in the mixed state if \(\alpha_{ab,2}>0\) is sufficiently large.

The GL model is simple enough that we can determine some of the conditions for the finite-momentum pairing to appear analytically. First, we note that \(\mathcal{F}\) always has at most three minima: at \(q=0\) or at \(q=\pm q_0\) (see Appendix \ref{AppC}). We also observe that \(\delta\mathcal{F}/\delta q\) is always zero at \(q=0\), so it is always a saddle point. This makes a first order phase transition possible across which \(q=0\) remains a local minimum solution while the \(q=q_0\) becomes the global minimum solution, see Fig. 1(e).
This is in contrast to the anapole model, in which \(\alpha_{ab}\) is linear in \(q\) to leading order and a kink occurs at \(q=0\) as a result, which makes the first order phase transition unlikely.

Next, optimizing \(\mathcal{F}\) over \(q\) at fixed \(|\Delta_a|\) and \(|\Delta_b|\) we have
\[q_0^2=-\frac{\alpha_{a,2}r^2-\alpha_{ab,2}r+\alpha_{b,2}}{2\left(\alpha_{a,4}r^2+\alpha_{b,4}\right)}=-\frac{C(r)}{2\left(\alpha_{a,4}r^2+\alpha_{b,4}\right)}\label{Eq:q0}\]
where \(r=|\Delta_a/\Delta_b|\).
For \(q_0\) to exist \(C(r)\) must be negative.
Note that \(C(r)=R\) with fixed \(R\) defines a conic section in the \((|\Delta_a|, |\Delta_b|)\), with discriminant \(D=\alpha_{ab,2}^2-4\alpha_{a,2}\alpha_{b,2}\) \footnote{Recall that if \(D<0\), \(C=R\) defines an ellipse and \(C\sim\tilde{x}^2+\tilde{y}^2\geq0\). If \(D>0\), \(C=R\) defines hyperbolas, and \(C\sim\tilde{x}^2-\tilde{y}^2\) can change sign.}.
\(C=0\) in particular defines two lines, \(|\Delta_a|=r_c(1\pm\sqrt{1-d})|\Delta_b|\), across which \(C\) changes sign, with \(r_c=\alpha_{ab,2}/(2\alpha_{a,2})\) and \(d=4\alpha_{a,2}\alpha_{b,2}/\alpha_{ab,2}^2\).
For \(C(r)\) to be negative, the gap ratio \(r\equiv|\Delta_a|/|\Delta_b|\) must be in the range
\(r\in r_c\left[1-\sqrt{1-d}, 1+\sqrt{1-d}\right]\).\footnote{We also note that there is an upper bound on \(q_0\) that can be obtained by solving for Eq. (\ref{Eq:q0}) for \(r(q_0)\): real solutions exist only if \(4 (\alpha_{a,2} + 2 \alpha_{a,4} q_0^2) (\alpha_{b,2} + 2 \alpha_{b,4} q_0^2) <  \alpha_{ab,2}^2\) (this can be easily reformulated as an explicit condition for \(q_0\), but the expression is a bit lengthy; for large \(\alpha_{ab,2}\), the condition is simply \(q_0^2<\alpha_{ab,2}/(4\sqrt{\alpha_{a,4}\alpha_{b,4}})\)).}
In particular, it is necessary that \(d<1\), which yields a necessary condition on the GL coefficients for SDE to exist: \(\alpha_{ab,2}^2>4\alpha_{a,2}\alpha_{b,2}\).

At a second order phase transition (if it occurs) inside the coexistence region, the gap ratio is the same as in the uniform \(q=0\) state, \(r=r_0=\sqrt{\alpha_{a,0}\beta_b/(\alpha_{b,0}\beta_a})\).
One can also check that the \(q=0\) saddle point becomes a maximum at the same time, since \(\delta^2\mathcal{F}/\delta q^2|_{q=0}=2C|\Delta_b|^2\), and consequently the \(q=0\) solution becomes a local maximum when \(C\) changes sign from positive to negative, which happens precisely when \(r=r_0=r_c(1\pm\sqrt{1-d})\equiv r_{c\pm}\).
As we find below in our numerical study, however, a first order phase transitions generically happen when \(\alpha_{ab,2}^2\gg4\alpha_{a,2}\alpha_{b,2}\), in which case the \(q=0\) saddle point remains a local minimum at the transition.
\\

\begin{figure*}[t]
\centering
\includegraphics[width=0.99\textwidth]{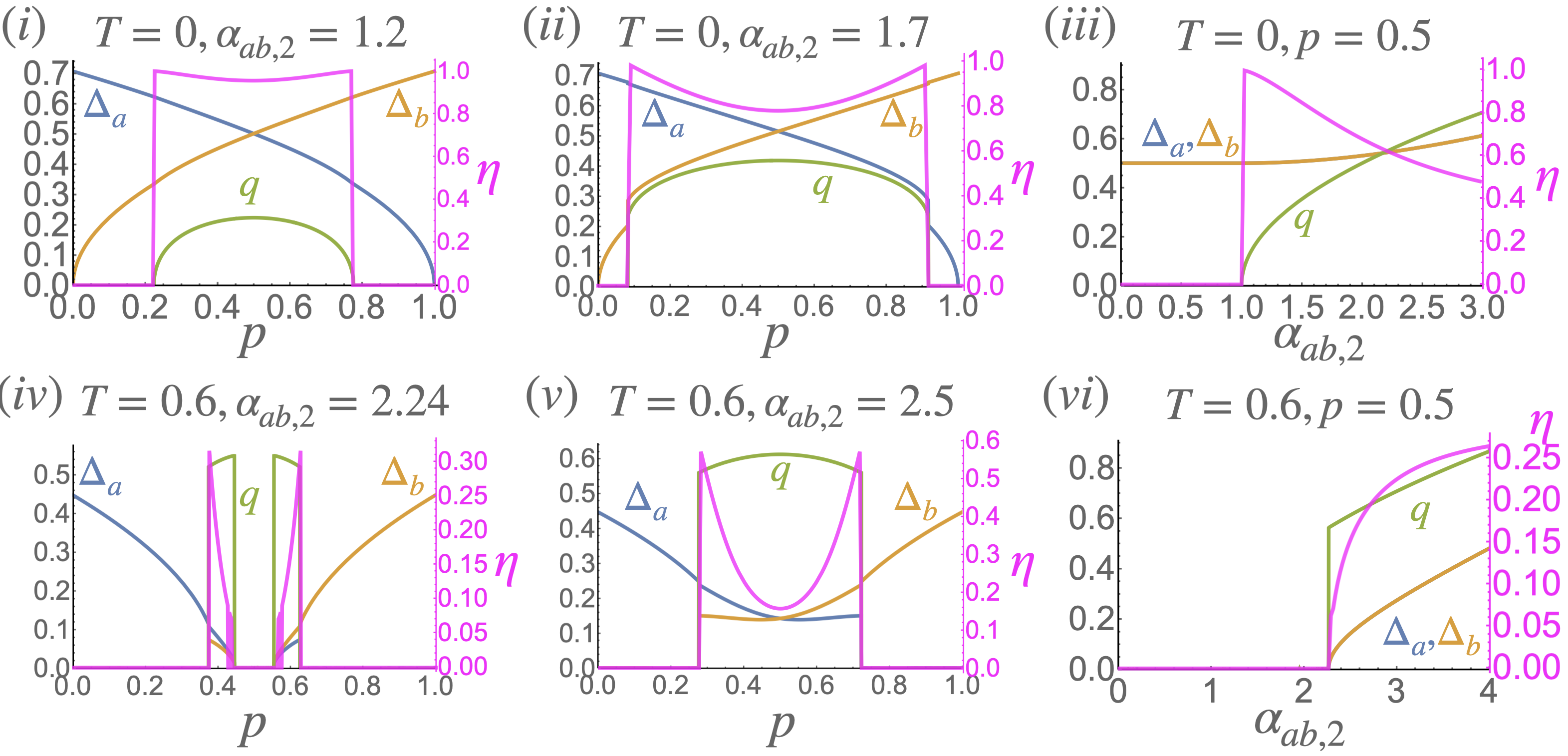}
\caption{Order parameters \(\Delta_a\) (blue), \(\Delta_b\) (orange) and \(q\) (green) computed across various phase transitions along cuts (i-vi) indicated in Fig. \ref{fig:PhaseDiagram}(b), with numerical values in arbitrary units on the left. The corresponding superconducting diode coefficient \(\eta\) is shown in magenta with numerical values on the right axis.}
\label{fig:OrderParameters}
\end{figure*}


\section{Phase Diagram and SDE}

We next study the phase diagram of the free energy Eq. \ref{EqGL} numerically, treating \(T\), \(p\) and \(\alpha_{ab,2}\) as variable parameters and taking \(\alpha_{f,2}=0.5\), \(\alpha_{f,4}=1\) and \(\beta_{f}=1\) for both \(f=a,b\) (such that the phase diagram is symmetric with respect to \(p\rightarrow1-p\)). The results are shown in Fig. \ref{fig:PhaseDiagram}. 
In Fig. \ref{fig:PhaseDiagram}(a) the red/blue surface separates the uniform (\(q=0\)) and nonuniform (\(q\neq 0\)) phases. \(\Delta_a\) and \(\Delta_b\) always coexist in the nonuniform phase, and coexist in the uniform phase when both \(T<T_{ca}(p)\) and \(T<T_{cb}(p)\) (shown as gray surface in Fig. \ref{fig:PhaseDiagram}(a) and gray dashed lines in Fig. \ref{fig:PhaseDiagram}(c)). See also Fig. \ref{fig:OrderParameters} where \(\Delta_a, \Delta_b\), and \(q\) are shown in blue, orange and green, respectively, for various fixed temperature cuts of the phase diagram indicated in Fig. \ref{fig:PhaseDiagram}(b). Fig. \ref{fig:PhaseDiagram}(d-f) additionally shows the condensation energy \(F(q)\) (the minimum of \(\mathcal{F}\) at fixed \(q\)) and the corresponding supercurrent \(J(q)=2\partial_q F(q)\) at various points in the phase diagram.

The nonuniform phase occurs only for \(\alpha_{ab,2}^2>4\alpha_{a,2}\alpha_{b,2}=1\).
For \(\alpha_{ab,2}\) not too far above that value the nonuniform phase occurs mostly in the coexistence region of the uniform phase (with \(T<\min[T_{ca}(p),T_{cb}(p)]\)). Close to the degeneracy point and for larger \(\alpha_{ab,2}\), however, the nonuniform phase appears outside of the uniform coexistence region, even including the region where no SC exists for \(\alpha_{ab,2}<1\) (namely when \(T>\max[T_{ca}(p),T_{cb}(p)]\). Fig. \ref{fig:PhaseDiagram}(b) shows the typical phase boundaries at fixed \(T\) below (solid line) and above (dashed line) the uniform degeneracy point at \(T=0.5\). For \(T<0.5\), at lower \(\alpha_{ab,2}\) the phase transition is second order: the minimum at \(q=0\) becomes unstable at the transition and the two minima at \(q=\pm q_0\) continuously emerge (see Fig. \ref{fig:PhaseDiagram}(d)). Below the transition in the uniform phase with \(q=0\), the critical currents \(J_{c+}\) and \(J_{c-}\) occur at \(q=q_{c+}>0\) and \(q=q_{c-}=-q_{c+}<0\). In the non-uniform phase, while \(J_{c+}\) still occurs at a slightly shifted \(q_{c+}>0\), the lower critical current corresponds to a new minimum of \(J(q)\) at \(q_{c-}>0\). Approaching the phase transition from above, \(J_{c-}\) therefore approaches zero, and the corresponding superconducting diode coefficient  \(\eta=(J_{c+}+J_{c-})/(J_{c+}-J_{c-})\) 
(with \(J_{c-}<0\)) therefore approaches its maximum possible value of \(1\), technically resulting in a ``perfect'' diode effect. This can be seen in Fig. \ref{fig:OrderParameters}(i) and (iii), where \(\eta\) is shown in magenta. A similar ``perfect'' diode effect was identified at the tricritical point of the FFLO state \cite{YuanFu22}. We note, however, and as noted in the supplementary in \cite{YuanFu22}, that \(\eta\) likely does not reach the maximal value in real systems as the nonuniform state can jump from the \(q_0\) to \(-q_0\) state before the transition happens, as the energy barrier for that jump vanishes at the transition.

At larger \(\alpha_{ab,2}\), the transition is similar for \(T\) either above or below the uniform degeneracy point \(T=0.5\) and becomes first order (see Fig. \ref{fig:PhaseDiagram}(e)): the minimum at \(q_0\) becomes a global minimum before the saddle-point at \(q=0\) switches from a minimum to a maximum. In this case \(q\), as well as \(\Delta_a\) and \(\Delta_b\), jump by a finite value at the phase transition, as shown in Fig. \ref{fig:OrderParameters}(ii) and (v), as well as the transitions in (iv) at the smallest and largest \(p\) values. In this case both \(J_{c+}\) and \(J_{c-}\) jump at the transition, so that \(\eta\) is in general not equal to \(1\) at the transition, but it is also not equal to zero and in general increases as the transition is approached from the 
nonuniform side.

Finally, a different kind of second order phase transition happens for \(T>T_{ca},T_{cb}\), as shown in Fig. \ref{fig:PhaseDiagram}(e): in this case the uniform solution does not exist, so no phase transition can occur between the uniform and nonuniform states. The order parameter \(\Delta_a\) and \(\Delta_b\), however, still vanish at the transition into the normal state (see Fig. \ref{fig:OrderParameters}(iv) and the phase transitions closest to \(p=0.5\) in (iv)). Consequently, while \(q\) still jumps by a finite value the superconducting diode coefficient \(\eta\) vanishes at this transition.
\\


\section{Microscopic derivation of GL coefficients}

In order for SDE to exist the GL coefficients must satisfy the necessary condition \(\alpha_{ab,2}^2>4\alpha_{a,2}\alpha_{b,2}\), as explained above. It is reasonable to ask whether this is a realistic condition from a microscopic perspective. Here we show that this it is indeed possible in a reasonable microscopic toy model, namely a two-band (or two-flavor) model with normal state Hamiltonian \(\mathcal{H}=\sum_{\mathbf{k},f}\left(\xi(\mathbf{k})c^\dagger_{\mathbf{k}f}c_{\mathbf{k}f}+t c^\dagger_{\mathbf{k}f}c_{\mathbf{k}\bar{f}}\right)\) where \(c_{\mathbf{k}f}\) annihilate electrons with momentum \(\mathbf{k}\) on band \(f=a,b\), and \(\bar{f}=a,b\neq f\). Assuming that the gap functions \(\hat{\Delta}_f\sim \langle c_{\mathbf{k}f}c_{-\mathbf{k+q}f}\rangle\), that there is no interband pairing, and taking for simplicity \(\hat{\Delta}_a=\Delta_a\) and \(\hat{\Delta}_b= \Delta_b \cos2\theta\), we find that in the weak-coupling approximation (with \(\xi(\mathbf{k})=\xi_{0}+\mathbf{v}_{F}\cdot \mathbf{k}\) linearized around the Fermi momentum)
\[\frac{\alpha_{ab,2}^2}{4\alpha_{a,2}\alpha_{b,2}}=\frac{1}{2} \left(\frac{14\zeta(3)+\text{Re}[\psi^{(2)}(1/2+it/(2\pi T))]}{14\zeta(3)-\text{Re}[\psi^{(2)}(1/2+it/(2\pi T))]}\right)^2\]
(see Appendix \ref{AppD} for calculation details). The function on the RHS takes values in the range \([0,1.1665]\), being zero at \(t=0\), peaking at \(t=\pi T\), and tending to \(0.5\) as \(t\rightarrow\infty\). 
There is therefore a range of \(t\) close to \(\pi T\) where the SDE indeed takes place in the phase diagram.
It is then reasonable to assume that the condition \(\alpha_{ab,2}^2>4\alpha_{a,2}\alpha_{b,2}\) for SDE can generically be realized in more sophisticated models relevant to real materials.
\\

\section{Discussion}

From a materials perspective, the possibility of SDE in multiphase SCs with equal parity mixing proposed in this article greatly extends the possibilities for searching for SDE. Though rare, there is now evidence for many multiphase SCs, including UPt\(_3\) \cite{JoyntTaillefer02}, UBe\(_{13}\) \cite{OttSmith85}, UTe\(_2\)
\cite{ShafferChichinadze22, WuChichinadzeShafferEaton23, YuAgterbergRaghu23, Machida23, WuEatonGrosche24, TeiFujimoto24, AokiYanase21, LewinButch23}, Sr\(_2\)RuO\(_4\) \cite{Agterberg21,GhoshRamshaw21}, Cd\(_2\)Re\(_2\)O\(_7\) \cite{KobayashiMurata11}, and PrOs\(_4\)Sb\(_{12}\) \cite{FrederickMaple07}.

From our microscopic two-band model we further learned that multiband SCs with dominant intraband pairing may favor the SDE. This suggests that multiband and multicomponent SCs including high \(T_c\) \cite{Sigrist98, FernandesChubukov16} but especially those that may favor intraband pairing -- such as locally \cite{YoshidaSigristYanase15, FischerSigristAgterbergYanase23} and globally \cite{BauerSigrist12, SmidmanAgterberg17} noncentrosymmetric, sublattice-polarized \cite{KieselThomale12, WuThomaleRaghu22, CastroShafferWuSantos23}, and (twisted moir\'{e} or untwisted) multilayer superconductors \cite{Liu17, MockliYanaseSigrsit18, KoblischkaKoblischka24, NuckollsYazdani24} -- are particularly promising candidates for realizing SDE.

In conclusion we note that given the high diode efficiencies that we find in the vicinity of the phase transitions where the SDE can be turned on and off (i.e. an implementation of a SC transistor) by small changes in external parameters without leaving the SC regime, such multiphase SC-based superconducting diodes may be particularly suitable for future technological applications.

\section*{Acknowledgements}

We thank Z. Wu, T. I. Weinberger, A. J. Hickey, and A. G. Eaton for useful discussions. This work was financially supported by the National Science Foundation, Quantum Leap Challenge Institute for Hybrid Quantum Architectures and Networks Grant No. OMA-2016136 (D.S. and A.L.). A. L. gratefully acknowledges H. I. Romnes Faculty Fellowship provided by the University of Wisconsin-Madison Office of the Vice Chancellor for Research and Graduate Education with funding from the Wisconsin Alumni Research Foundation. D.V.C. acknowledges financial support from the National High Magnetic Field Laboratory through a Dirac Fellowship, which is funded by the National Science Foundation (Grant No. DMR-1644779) and the State of Florida.
This paper was written in-part at the Aspen Center for Physics, during the program ``Quantum Matter Through the Lens of Moir\'e Materials", which is supported by the NSF Grant No. PHY-2210452. 

\appendix

\section{Ginzburg-Landau Saddle Point Equations for Mixed-Parity Case}\label{AppA}

For the case of mixed-parity mixing considered, e.g., for the case of anapole SC in \cite{KanasugiYanase22}, \(\alpha_{ab}(q)\) is an odd function of \(q\). To leading order in \(q\), we can thus take \(|\alpha_{ab}|=\alpha_{ab,1}|q|\) with \(\alpha_{ab,1}\geq0\). The saddle-point equations are then
\begin{align}
    \frac{\delta\mathcal{F}}{\delta|\Delta_a|}&=2\alpha_a=|\Delta_a|-2\alpha_{ab,1}q|\Delta_b|+4\beta_a|\Delta_a|^3=0\nonumber\\
    \frac{\delta\mathcal{F}}{\delta|\Delta_b|}&=2\alpha_b|\Delta_b|-2\alpha_{ab,1}q|\Delta_a|+4\beta_b|\Delta_b|^3=0\nonumber\\
    \frac{\delta\mathcal{F}}{\delta q}&=2(\alpha_{a,2}|\Delta_a|^2+\alpha_{b,2}|\Delta_b|^2)q-2\alpha_{ab,1}|\Delta_a||\Delta_b|=0\nonumber
\end{align}
The last equation implies that in the ground state
\[q_0=\frac{\alpha_{ab,1}|\Delta_a||\Delta_b|}{\alpha_{a,2}|\Delta_a|^2+\alpha_{b,2}|\Delta_b|^2}\,;\label{q0}\]
note that \(\Delta_{a/b}\) are themselves functions of \(q\) in general, but we can see immediately that \(q_0\neq0\) iff \(|\Delta_a||\Delta_b|\neq 0\), i.e. the mixed state is always non-uniform. Moreover, at \(q=0\) we have
\[\left.\frac{\delta^2\mathcal{F}}{\delta q^2}\right|_{q=0}=2(\alpha_{a,2}|\Delta_a|^2+\alpha_{b,2}|\Delta_b|^2)>0\]
but \(\delta\mathcal{F}/\delta q|_{q=0}=-2\alpha_{ab,1}|\Delta_a||\Delta_b|\leq 0\). This means that the \(q=0\) state ceases to be a minimum in the mixed state whenever \(\alpha_{ab,1}\neq0\). In particular, this means that no first order phase transitions are possible in the mixed-parity case.

\section{Additional Properties of Saddle Points of Ginzburg-Landau Free Energy}

After some algebra, the equation for the mixed state at fixed \(q\) (i.e. the first equation in Eq. (2)) turns out to be simplest to express in terms of \(x_{a/b}=2\beta_{a/b}(|\Delta_{a/b}|^2-|\Delta_{a0/b0}|^2)\) where \(|\Delta_{a0/b0}|^2=-\alpha_{a/b}/(2\beta_{a/b})\). The equations become
\begin{align}
    &x_a x_b=\frac{\alpha_{ab,2}^2q^4}{4}=\frac{\alpha_{ab}^2}{4}\\
    &16 \frac{\beta_b}{\beta_a}x_a^3(x_a-\alpha_a)+4\alpha_b\alpha_{ab}^2 x_a=\alpha_{ab}^4\label{xaEq}
\end{align}
Note that \(x_a-\alpha_a\geq0\) and \(x_b-\alpha_b\geq0\) (another notable relation at the saddle point is \(|\Delta_a/\Delta_b|^2=x_b/x_a\)). A non-trivial consequence of the saddle-point equations is that \(x_a,x_b\geq0\) at the saddle points. As a consequence of that, there is always a single unique solution for \(x_a\): the LHS of Eq. (\ref{xaEq}) can only change sign once at some \(x_a\) (and only if \(\alpha_b<0\)), and it is positive and monotonic for \(x_a\) above that value. To find the critical current, we need to plug in the optimal solution into \(\mathcal{F}\) to obtain the condensation energy \(F(q)\). The current is then given by \(J(q)=2\partial_q F(q)\). We do this numerically to obtain the plots in Fig. 1 (d-f).

\section{Additional Symmetry-Allowed Quartic Terms in Ginzburg-Landau Free Energy}\label{AppC}

As mentioned in the main text, there are additional symmetry-allowed quartic terms that we have neglected in our model, for example \(\beta_{ab}|\Delta_a|^2|\Delta_b|^2\), as well as \(\beta_{abbb}(q)\Delta_a|\Delta_b|^2\Delta_b^*\), etc. 
In general, the additional quartic terms can lead to a nontrivial evolution of the complex phase between \(\Delta_a\) and \(\Delta_b\), and the analysis of the GL saddle point equations become less tractable, but the qualitative picture is not altered, as SDE still emerges in the mixed state. In particular, note that the \(\alpha_{ab,2}\) term is more relevant than the quartic terms, since it always dominates close enough to \(T_c\).

\begin{figure*}[t]
\centering
\includegraphics[width=0.99\textwidth]{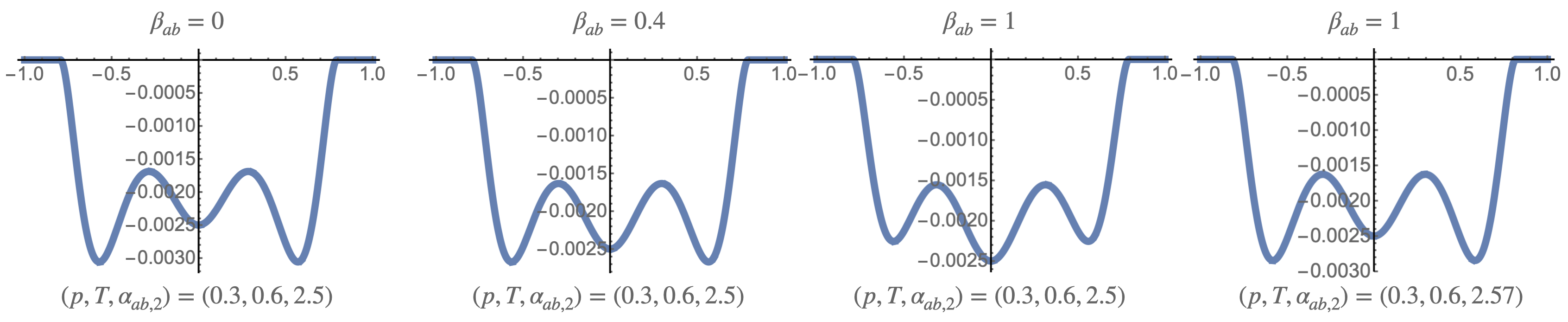}
\caption{Free energy \(\mathcal{F}(q)\) for parameter values around those in Fig. \ref{fig:PhaseDiagram}(e), with added quartic terms \(\beta_{ab}\).}
\label{fig:QuarticTerms}
\end{figure*}

We verify this numerically by adding the quartic term \(\beta_{ab}\), which only slightly changes the critical value of \(\alpha_{ab,2}\) above which the SDE takes place. The change can go in either direction depending on the values of the added terms, see Fig. \ref{fig:QuarticTerms} for parameters corresponding roughly to Fig. \ref{fig:PhaseDiagram}(e). Increasing \(\beta_{ab}\) in this case takes the system from the phase with SDE to the phase without SDE, but a small adjustment of \(\alpha_{ab}\) brings the global minimum back to the finite \(q\) phase with SDE.

\section{Details of Derivation of GL Coefficients from Microscopic Model}\label{AppD}

\begin{figure}[t]
\centering
\includegraphics[width=0.49\textwidth]{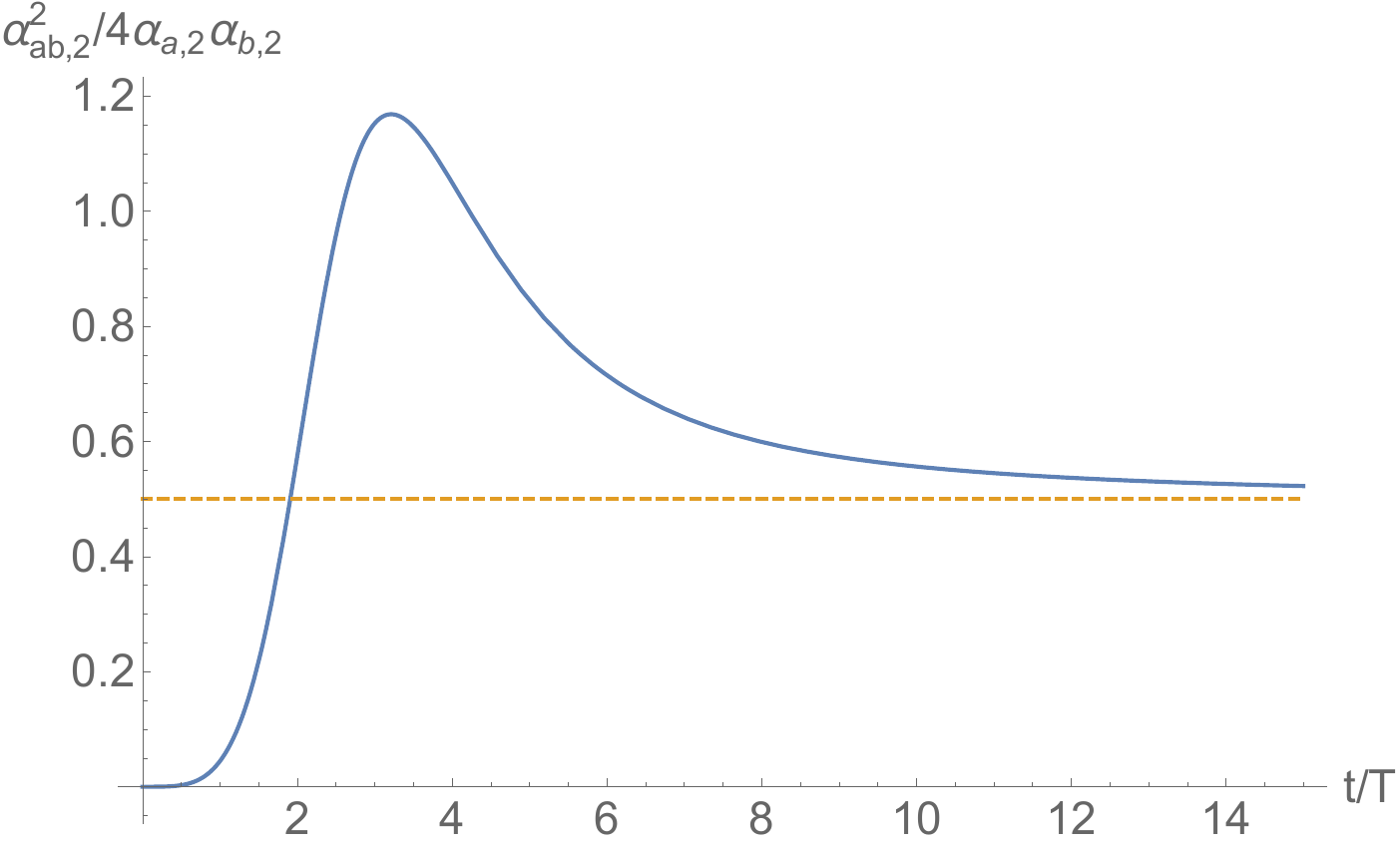}
\caption{Plot of the dimensionless ratio of GL coefficients defined by Eq. \eqref{eq:alpha-ratio}.}
\label{fig:alpha-ratio}
\end{figure}

Here we derive the coefficients \(\alpha_{f,2}\) and \(\alpha_{ab,2}\) for the two-band microscopic toy model in the main text with Hamiltonian
\[\mathcal{H}=\sum_{\mathbf{k},f}\left(\xi(\mathbf{k})c^\dagger_{\mathbf{k}f}c_{\mathbf{k}f}+t c^\dagger_{\mathbf{k}f}c_{\mathbf{k}\bar{f}}\right)\]
The Ginzburg-Landau free energy is obtained in the standard way by introducing the Hubbard-Stratonovich fields \(\widehat{\Delta}_f\) which enter the transformed Hamiltonian as \(\widehat{\Delta}_f(\mathbf{p;q})c^\dagger_{\mathbf{k+q}/2,f}c^\dagger_{-\mathbf{k+q}/2,f}\) (and we assume no interband pairing between \(a\) and \(b\) bands):
\begin{widetext}
\[\mathcal{F}[\widehat{\Delta},\widehat{\Delta}^\dagger]=-T\sum_{n\mathbf{k} j} \frac{1}{2j}\text{Tr}\left[\left(\widehat{\Delta}^\dagger(\mathbf{k;q})G^{(0)}(i\omega_n,\mathbf{k+q}/2)\widehat{\Delta}(\mathbf{k;q})G^{(0,h)}(i\omega_n,\mathbf{k-q}/2)\right)^j\right]+H_{\Delta^2}\label{Fexpansion}\]
where 
\[\widehat{\Delta}=\left(\begin{array}{cc}
   \widehat{\Delta}_a  & 0 \\
    0 & \widehat{\Delta}_b
\end{array}\right)\]
and
\[H_{\Delta^2}=\sum_{\substack{\mathbf{p,k}\\ }}\widehat{\Delta}^\dagger_{f_3f_4}(\mathbf{k;q})\left[V^{-1}(\mathbf{p,k;q})\right]^{f_3f_4}_{f_1f_2}\widehat{\Delta}_{f_1f_2}(\mathbf{p;q})\,.\]
Here \(V(\mathbf{p,k;q})\) are the pairing interactions and \(f_j\) are band indices (summed over implicitly). 
Assuming the interactions do not depend on the total Cooper pair momentum \(q\) and that the gap functions are self-consistent, the free energy to second order in \(\Delta\) is
\begin{align}
\mathcal{F}_2&=\frac{T}{2}\sum_{n\mathbf{k}}\text{Tr}\left[\widehat{\Delta}^\dagger G^{(0)}(i\omega_n,\mathbf{k+q/2})\widehat{\Delta}G^{(0,h)}(i\omega_n,\mathbf{k-q/2})\right]-\frac{T_c(0)}{2}\sum_{n\mathbf{k}} \text{Tr}\left[\widehat{\Delta}^\dagger G^{(0)}(i\omega_n|_{T_c},\mathbf{k})\widehat{\Delta}G^{(0,h)}(i\omega_n|_{T_c},\mathbf{k})\right]
\end{align}
where \(G^{(0)}\) is the Green's function which is \(2\times2\) in the sublattice indices, \(G^{(0,h)}(i\omega,\mathbf{k})=-G^T(-i\omega,-\mathbf{k})\).
In the simplified model,
\[
    G^{(0)}(i\omega_n,\mathbf{k})=\left(\begin{array}{cc}
    G_{1}(i\omega_n,\mathbf{k}) & G_{2}(i\omega_n,\mathbf{k})  \\
    G_{2}(i\omega_n,\mathbf{k}) & G_{1}(i\omega_n,\mathbf{k})
\end{array}\right)=\frac{1}{( i\omega_n-\xi(\mathbf{k}))^2-t^2}\left(\begin{array}{cc}
    i\omega_n-\xi(\mathbf{k}) & t  \\
    t & i\omega_n-\xi(\mathbf{k}) 
\end{array}\right)\]
Taking \(\xi(\mathbf{k})=\xi_{0}+\mathbf{v}_{F}\cdot \mathbf{k}\) and \(\widehat{\Delta}_f(\mathbf{k;q})=\Delta_f d_f(\mathbf{k})\), we have
\[\mathcal{F}_2=\left\langle \Pi_1(\theta,\mathbf{q})\left(d_a^2(\mathbf{k})|\Delta_a|^2+d_b^2(\mathbf{k})|\Delta_b|^2\right)+2\Pi_2(\theta,\mathbf{q})d_a(\mathbf{k})d_b(\mathbf{k})|\Delta_a||\Delta_b|\right\rangle_{\theta}\]
where
\begin{align}
    \Pi_1(\theta,\mathbf{q})&=N\left(-\log\frac{T_c}{T}+\digamma_1\left(\delta \xi(\theta,\mathbf{q})/T,t/T\right)\right)\nonumber\\
    \Pi_2(\theta,\mathbf{q})&=N\digamma_2\left(\delta \xi(\theta,\mathbf{q})/T,t/T\right)
\end{align}
where \(N\) is the DOS, \(\delta \xi(\theta,\mathbf{q})=\mathbf{v}_{F}(\theta)\cdot \mathbf{q}/2=(v_{Fx}\cos\theta q_x+v_{Fy}\sin\theta q_x)/2\), and
\begin{align}
    \digamma_1\left(x,y\right)&=\frac{1}{4}\text{Re}\left[2\psi\left(\frac{1}{2}+\frac{i x}{2\pi}\right)+\psi\left(\frac{1}{2}+\frac{i (x+y)}{2\pi}\right)+\psi\left(\frac{1}{2}+\frac{i (x-y)}{2\pi}\right)-2\psi\left(\frac{1}{2}+\frac{i y}{2\pi}\right)-2\psi\left(\frac{1}{2}\right)\right]\nonumber\\
    \digamma_2\left(x,y\right)&=\frac{1}{4}\text{Re}\left[2\psi\left(\frac{1}{2}+\frac{i x}{2\pi}\right)-\psi\left(\frac{1}{2}+\frac{i (x+y)}{2\pi}\right)-\psi\left(\frac{1}{2}-\frac{i (x-y)}{2\pi}\right)+2\psi\left(\frac{1}{2}+\frac{i y}{2\pi}\right)-2\psi\left(\frac{1}{2}\right)\right]\label{digamma}
\end{align}
and \(\langle\dots\rangle_\theta=\int_0^{2\pi} \dots d\theta/(2\pi)\), with \(\theta\) the angle around the FS (we assume everything can be considered as a function of this angle and is thus independent of \(|\mathbf{k}|\)).

If we now take \(d_a=1\) and \(d_b=\cos2\theta\) for sake of argument (i.e. \(s\)-wave and \(d\)-wave pairing) and expand in \(q\), we get
\begin{align}
    \alpha_{a}(\mathbf{q})&=\alpha_{a,0}+\frac{14\zeta(3)-\text{Re}[\psi^{(2)}(1/2+it/(2\pi T))]}{64\pi^2T^2}N(v_{Fx}^2q_x^2+v_{Fy}^2q_y^2)+\dots\nonumber\\
    \alpha_{b}(\mathbf{q})&=\alpha_{b,0}+\frac{14\zeta(3)-\text{Re}[\psi^{(2)}(1/2+it/(2\pi T))]}{128\pi^2T^2}N(v_{Fx}^2q_x^2+v_{Fy}^2q_y^2)+\dots\nonumber\\
    \alpha_{ab}(\mathbf{q})&=\frac{14\zeta(3)+\text{Re}[\psi^{(2)}(1/2+it/(2\pi T))]}{64\pi^2T^2}N(v^2_{Fx}q_x^2-v_{Fy}^2q_y^2)+\dots
\end{align}
\end{widetext}
We thus have
\begin{align}
    \frac{\alpha_{ab,2}^2}{4\alpha_{a,2}\alpha_{b,2}}&\propto \left(\frac{14\zeta(3)+\text{Re}[\psi^{(2)}(1/2+it/(2\pi T))]}{14\zeta(3)-\text{Re}[\psi^{(2)}(1/2+it/(2\pi T))]}\right)^2
\end{align}
(the ratio is zero at \(t=0\), maximized for \(t=\pi T\), and tends to a constant value as $t/T\to\infty$). Assuming \(v_{Fx}=v_{Fy}\), the free energy is maximized for \(\mathbf{q}\) along \(x\) or \(y\) directions, and
\begin{align}\label{eq:alpha-ratio}
\frac{\alpha_{ab,2}^2}{4\alpha_{a,2}\alpha_{b,2}}&=\frac{1}{2} \left(\frac{14\zeta(3)+\text{Re}[\psi^{(2)}(1/2+it/(2\pi T))]}{14\zeta(3)-\text{Re}[\psi^{(2)}(1/2+it/(2\pi T))]}\right)^2,
\end{align}
$\in [0,1.1665]$ see Fig. \ref{fig:alpha-ratio}.
There is therefore a range of \(t\) (close to \(\pi T\)) for which the mixed phase does appear, and it is not unreasonable to expect even larger values of \(\frac{\alpha_{ab,2}^2}{4\alpha_{a,2}\alpha_{b,2}}\) in more sophisticated models.

\bibliography{SCdiode}

\end{document}